\documentclass[11pt]{article}
\usepackage{times}
\usepackage[left=2.2cm, right=2.2cm, top=2cm, bottom=2.5cm]{geometry}

\usepackage[utf8]{inputenc}
\usepackage{enumitem,amssymb}
\usepackage{ragged2e}
\newlist{thematic}{itemize}{8}
\setlist[thematic]{label=$\square$}
\usepackage{pifont}
\usepackage{fancyhdr}
\usepackage[USenglish]{babel}
\usepackage[utf8]{inputenc}
\usepackage[T1]{fontenc}
\usepackage{epsfig}
\usepackage{wrapfig}
\usepackage{color}
\usepackage[most]{tcolorbox}

\usepackage[vflt]{floatflt}
\usepackage{longtable}
\usepackage{caption}
\usepackage{jheppub}   
\setcitestyle{authoryear,round}

\usepackage[dvipsnames,svgnames,x11names]{xcolor}
\usepackage[all]{hypcap} 
\usepackage{amssymb,amsmath}
\usepackage{longtable}
\usepackage{amssymb}
\usepackage{amsmath, xspace}
\usepackage{graphics,graphicx} 
\usepackage{rotating}
\usepackage{color}
\usepackage{xcolor}
\usepackage{multirow}
\usepackage{subfigure}
\usepackage{sectsty}
\usepackage[outercaption]{sidecap}
\sidecaptionvpos{figure}{c}

\newcommand{\nnhp}{$\rm N_2H^+$\xspace}
\newcommand{\nndp}{$\rm N_2D^+$\xspace}
\newcommand{\ohhdp}{ortho-$\rm H_2D^+$\xspace}
\newcommand{\pddhp}{para-$\rm D_2H^+$\xspace}
\newcommand{\hhdp}{$\rm H_2D^+$\xspace}
\newcommand{\ddhp}{$\rm D_2H^+$\xspace}

\usepackage{enumitem}
\setlist[enumerate]{itemsep=0pt, parsep=0pt}
\setlist[itemize]{itemsep=0pt, parsep=0pt}

\definecolor{DarkGreen}{rgb}{0.0, 0.3, 0.0}
\definecolor{purple}{rgb}{0.5, 0.0, 0.5}
\definecolor{red}{rgb}{1, 0.0, 0.0}
\definecolor{green}{rgb}{0, 1.0, 0.0}

\def\3he{$^3{\rm He}$}

\def\lsim{\mathrel{\lower2.5pt\vbox{\lineskip=0pt\baselineskip=0pt
           \hbox{$<$}\hbox{$\sim$}}}}

\def\gsim{\mathrel{\lower2.5pt\vbox{\lineskip=0pt\baselineskip=0pt
           \hbox{$>$}\hbox{$\sim$}}}}

\begin{document}
\raggedright
\huge
Chemical and dynamical evolution of prestellar cores, the dawn of star and planet formation\\  

\bigskip
\normalsize

\textbf{Authors:} 
Silvia Spezzano (spezzano@mpe.mpg.de, MPE);
Elena Redaelli (ESO); Paola Caselli (MPE); Doris Arzoumanian (Kyushu University), Arnaud Belloche (MPIfR), Doug Johnstone (NRC Canada), Tony Mroczkowski (ICE-CSIC) , \'Alvaro S\'anchez-Monge (ICE-CSIC), Olli Sipil\"a (MPE), Sigurd S. Jensen (MPE), Izaskun Jim\'enez-Serra (CAB).
\linebreak

\textbf{Science Keywords:} Galaxy: prestellar cores, star and planet formation. ISM: kinematics and dynamics.
\linebreak

 \captionsetup{labelformat=empty}
\begin{figure}[h]
   \centering
\includegraphics[width=.9\textwidth]{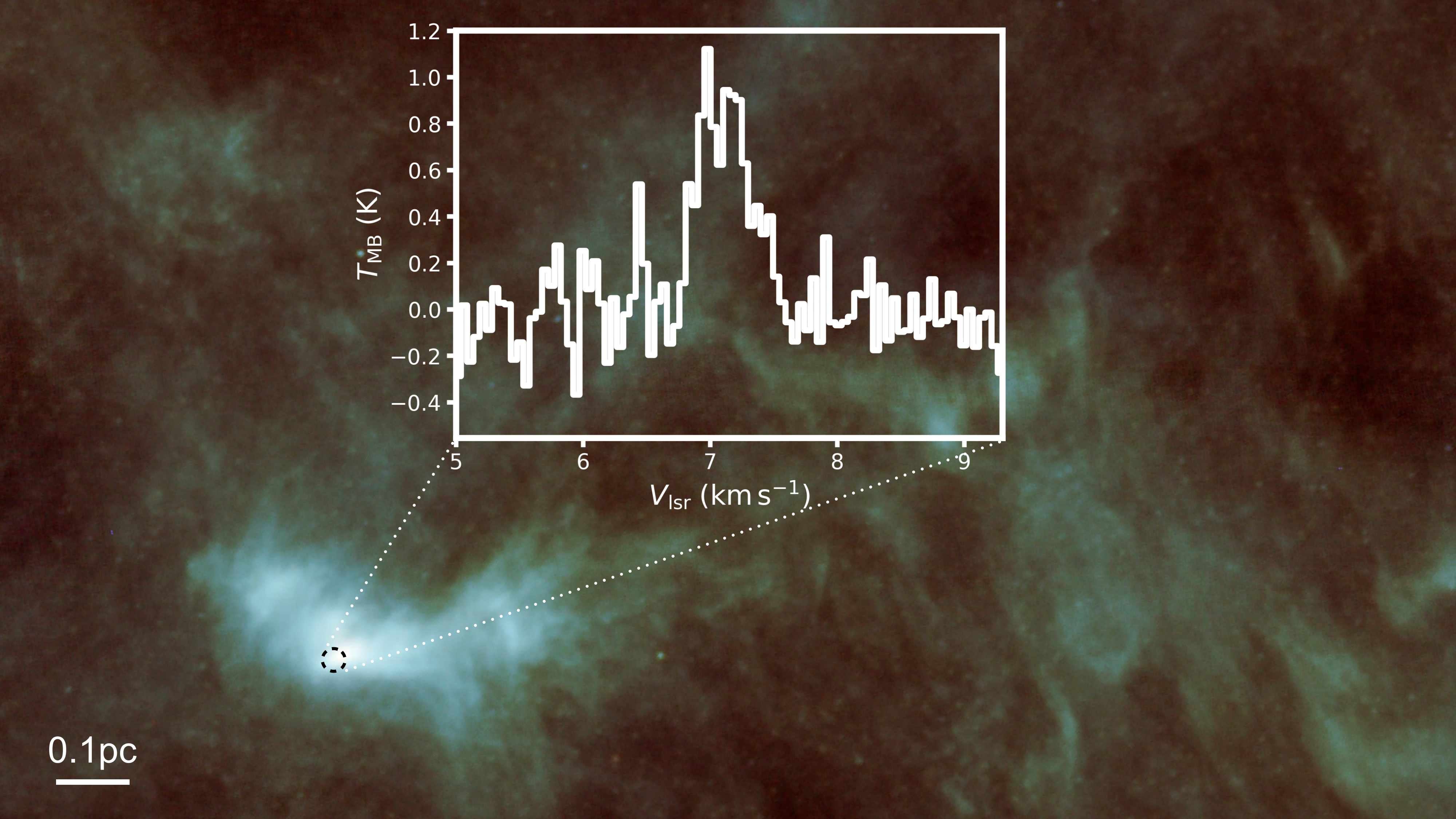}
   \caption{The dust thermal emission towards the prestellar core L1544 and its surroundings at 250$\,\mu$m, detected with the \textit{Herschel} space mission, is shown in colourscale. The inset shows the \ohhdp spectrum towards the core's centre as seen by APEX. Image credits: ESA/Herschel/SPIRE.
 }
\end{figure}
\vspace{-15mm}

\setcounter{figure}{0}
\captionsetup{labelformat=default}


\pagebreak
\justifying

\noindent \textbf{\Large Abstract:}  Studying prestellar cores is crucial to understanding the formation of stellar and planetary systems. To date, we lack a comprehensive survey of their density and velocity structure at small scales (100-1000\,au), the gas ionisation state, their lifetimes, and how these properties are influenced by the environment. A deep study into this critical stage of star formation requires a large single-dish (sub)millimetre telescope like AtLAST.
\vspace{-8pt}


 \vspace{-12pt}
\section{Scientific context and motivation}
\vspace{-8pt}
Stellar systems form from the collapse of dense cores within filamentary structures in molecular clouds. Therefore, dense cores represent the initial conditions for star and planet formation.
Prestellar cores are dynamically evolved starless cores on the verge of gravitational collapse \citep{crapsi05}. Because of their short lifetime (a few 10$^5$ years), prestellar cores are rare. 
 The material that will finally be inherited by the next stages in the process of star and planet formation is already present at the prestellar phase, as already shown for water and complex organic molecules \citep{cleeves14, scibelli21, vangelder20}. The chemical structure of prestellar cores, i.e. the distribution of different molecular tracers within and around them, is quite complex and has been shown to depend on the evolutionary stage as well as on the environment where they are embedded (e.g., \citealt{spezzano20,nagy19}). Studying the chemistry and physics within prestellar cores and in their nearby surroundings is crucial to understanding the process of star and planet formation, as well as the chemical budget of material finally inherited by planets.
 \vspace{-12pt}
\subsection{Deuteration as diagnostic tool}
\vspace{-8pt}
Isotopic fractionation is a powerful tool to follow the evolution of material throughout the star-formation process. Prestellar cores exhibit some of the highest levels of deuteration (D/H $\geq$0.1, \citealt{crapsi05, pagani07, spezzano25}) because their physical conditions (low temperatures, $\lesssim$ 10 K, and high densities, $\gtrsim$10$^5$ cm$^3$) greatly favour the deuteration processes. The deuteration of $\rm H_3^+$ becomes very efficient for two reasons: (i) CO, one of the main destruction partners of $\rm H_3^+$ and its deuterated forms, disappears from the gas phase due to catastrophic freeze-out onto the cold dust grains \citep{caselli99}; (ii) $\rm H_3^+$ deuteration reactions are exothermic and, at temperatures below 30$\,$K, they efficiently produce \hhdp, \ddhp and $\rm D_3^+$. The abundances of these species, hence, show a sharp increase. The deuterated forms of H$_3^+$ are the principal agents of deuteration in dense and cold gas. Additionally, their dissociative recombination increases the atomic D/H ratio, leading to efficient deuteration of molecules on the dust grains via the adsorption of atomic D \citep{caselli12}. By comparing observed molecular deuteration fractions with those predicted by models, we can probe both the chemical and the dynamical evolution of the observed core.

 \vspace{-12pt}
\subsection{Mapping molecular ions to unveil the initial conditions of star and planet formation}
\begin{figure}[ht]
    \centering
    \begin{minipage}[b]{0.55\textwidth}
        \includegraphics[width=\linewidth]{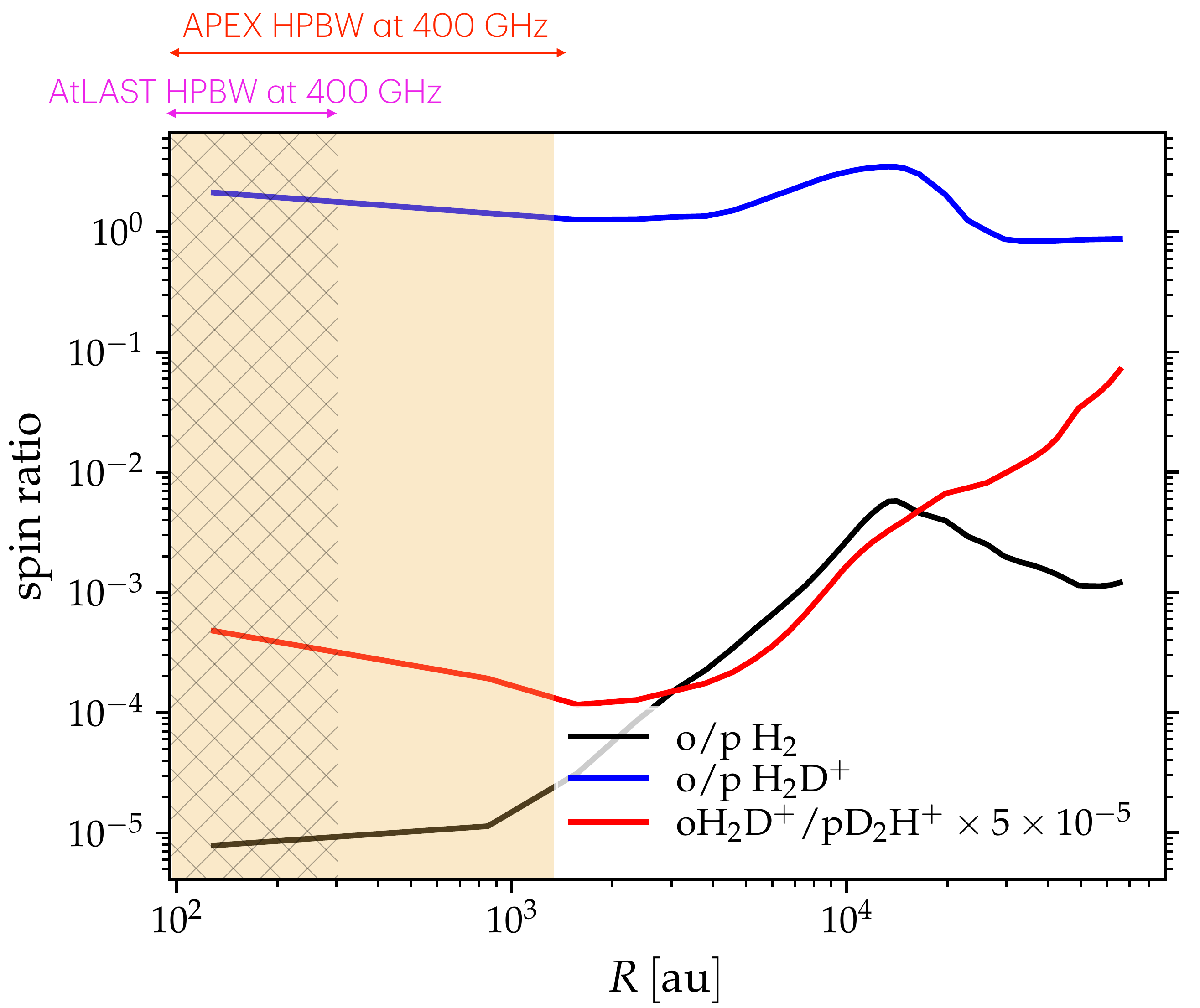} 
    \end{minipage}%
    \hfill
    \begin{minipage}[b]{0.38\textwidth}
        \caption{Radial profiles of the spin ratios of H$_2$, H$_2$D$^+$, and D$_2$H$^+$  computed using state-of the art chemical models \citep{sipila19} using the 1D physical structure of the prototypical prestellar core L1544 at the best fit time \citep{caselli22}. The model shows that the $o$H$_2$D$^+$/$p$D$_2$H$^+$ ratio is a good proxy for the o/p ratio in molecular hydrogen, H$_2$. The APEX and AtLAST HPBW at 400$\,$GHz, 2500 au (14$"$) and 600 au (3.5$"$) at the distance of L1544, are shown in the figure to highlight the importance of high angular resolution to resolve the core's structure and best compare with the chemical models.}
  \label{fig:spin_ratios}
    \end{minipage}
\end{figure}

 \vspace{-8pt}
One of the biggest advantages of using molecular emission to study astrophysical objects is that molecules are not homogeneously distributed within them. Beyond the differing spatial distributions of various molecules, distinct molecular transitions probe different layers of the gas, depending on their critical densities.\\
{\large \textcolor{blue}{N$_2$H$^+$ and N$_2$D$^+$:}}
The deuteration level of N$_2$H$^+$, a late-type molecule particularly abundant in the dense central regions of prestellar cores, is an excellent tool to estimate the dynamical evolution of dense cloud cores \citep{crapsi05, spezzano25, caselli25, redaelli25}. 
Additionally, mapping high-$J$ N$_2$H$^+$ and N$_2$D$^+$ transitions ($J$=5-4 and 6-5 at $\sim$400 GHz) is crucial to derive the spatial extent of the highly deuterated (N$_2$D$^+$/N$_2$H$^+$ $>$20$\%$) and very dense ($n_{H_2}$ $>$ few 10$^5$ cm$^{-3}$) central gas of prestellar cores. 
These lines have a high critical density ($\sim$10$^7$ cm$^{-3}$), and therefore their emission comes mainly from the densest gas at the centre of the prestellar core. By mapping this emission, we can probe the high density gas in a way that is independent of model deprojections typical in many areas of astrophysics.
This diagnostic power of N$_2$H$^+$ (5-4) and N$_2$D$^+$ (6-5) has been well demonstrated in \cite{spezzano25} and  \cite{redaelli25}.\\
{\large \textcolor{blue}{H$_2$D$^+$ and D$_2$H$^+$}}:
They are fundamental players in the evolution of dense cores and in shaping their chemical structure. In particular, they provide information about the electron fraction, a crucial parameter for the dynamical evolution of magnetized clouds (via coupling/decoupling of matter and magnetic fields) and have important effects on the abundance and deuteration of the large majority of other molecules also in dense cores. They can also be used to trace the abundance of $\rm H_3^+$, the starting point of gas-phase ion chemistry, which in dense gas cannot be directly observed due to the lack of detectable emission lines. \\
These species are also our only clue to the evolution of molecular hydrogen's ortho-to-para ratio (OPR). This is a key parameter that regulates the deuteration process, as the H$_2$ spin state impacts the exothermicity of the reaction producing \hhdp (see e.g., \citealt{pagani92}). The OPR is thus crucial to determine the timescales of dense core evolution (cf. Fig.~1). The ground state transitions of \ohhdp and \pddhp are the only ones that can be readily observed from the ground, at $0.8$ and $0.4\,$mm (372 and 692 GHz), respectively. The corresponding lines of para-$\rm H_2D^+$ and ortho-$\rm D_2H^+$ lie at $\sim 1.3-1.5\, $THz, are much more difficult to access due to the poor atmospheric transmission and technical difficulties. 
As shown in Fig.~\ref{fig:spin_ratios}, the \ohhdp/\pddhp ratio is a good indicator of the H$_2$ OPR, enabling us to investigate the lifespan of dense cores prior to their collapse.
\vspace{4pt}
\par
Given their \textit{\textbf{fundamental importance}} for the chemical development in dense cores, obtaining spatially resolved observations of the high-$J$ transitions of  N$_2$D$^+$ and N$_2$H$^+$ and the ground states of H$_2$D$^+$ and D$_2$H$^+$ will allow us to constrain chemical models significantly. So far, these detections (especially D$_2$H$^+$) remain quite elusive, especially because of the high frequencies.

\vspace{-12pt}
\section{Science goals}
 \vspace{-8pt}
In the context of what is described above, we can outline the following, pressing science questions:
\vspace{-4pt}
\begin{itemize}
\setlength{\itemindent}{-5pt}
    \item SG1: What is the \textbf{density and velocity structure} of prestellar cores at high spatial resolution (hundreds of au)? Constraining the physical properties of the dense gas in the center of prestellar cores will allow to test different models of collapse \citep{larson69, penston69, shu77} and to observe eventual fragmentation. To do so, it is paramount to use the appropriate tracers of high-density and cold molecular gas and observe them with high-velocity resolution ($\lesssim 0.05\,\rm km \, s^{-1}$). The ideal solution is to use several transitions of the nitrogen-bearing ion \nnhp and its deuterated counterpart \nndp at high angular resolution.
    \item SG2: What is the typical \textbf{ionisation degree} of dense cores?  This elusive parameter is quintessential to the dynamic evolution of star forming regions, as it sets the degree of coupling between matter and magnetic fields. It is, hence, a crucial input for any MHD simulation. \ohhdp and \pddhp are valuable tracers of the ionisation fraction of gas throughout the core, at intermediate/high densities.
    \item SG3: What is the typical \textbf{lifetime of the prestellar phase} embedded in different environments? 
    Prestellar cores are the first step in the formation of low-mass stars.
    It is therefore of greatest importance to measure their lifetime via OPR measurements as it affects their chemical and dynamical evolution.
\end{itemize}
\begin{tcolorbox}[colback=blue!5!white,colframe=blue!75!black]
To address these scientific questions, we need high resolution ($\sim 500\,$au) and high-frequency (up to $\sim 700 \,$GHz) observations that are also able to cover the large-scale emission ($\sim 10,000\,$au) of dense cores, providing the complete structure of the whole core, needed to constrain chemical and dynamical models. Furthermore, to study the initial conditions in the formation of low-mass stars in a statistical manner, and to assess environmental effects, a survey of cores is essential, hence requiring high sensitivity. These observational requirements are not met by any facility currently available or foreseen for the near future. 
\end{tcolorbox}
\vspace{-15pt}
\section{Scientific and Technical requirements}
\vspace{-8pt}
At present, we lack a statistically significant and unbiased survey to tackle the chemical and dynamical evolution of prestellar cores. Such a survey would establish a critical baseline for the study of star and planet formation, as well as the astrochemical evolution during these processes. To reach our goals and to address the scientific questions listed in the previous section, we require to map the following transitions in a sample of 50 prestellar cores: the $J$=5-4 and 6-5 transitions of N$_2$H$^+$ and N$_2$D$^+$ at 460$\,$GHz, and the ground state transitions of \ohhdp and \pddhp at 372 and 692$\,$GHz. 
The necessary technical requirements to obtain observations that will have the potential to revolutionize the current status of this research field are:
\vspace{-4pt}
\begin{enumerate}
\item{\textbf{Multi-line and multi-band observations} that cover the range from 0.4$\,$mm (740$\,$GHz) to 1$\,$mm (300$\,$GHz).}
\item{\textbf{High-frequency resolution} (50$\,$kHz, 0.05$\,\rm km \, s^{-1}$ at 300 GHz) to resolve the line profiles (typical FWHM $\lesssim 0.3 \, \rm km\, s^{-1}$) and use the mapped lines to trace the velocity structure of the cores at the different depths traced by different molecules and molecular transitions \citep{ferrer22}. We note that a higher frequency resolution (10$\,$kHz, 0.03$\,\rm km \, s^{-1}$ at 100 GHz), will be necessary for precision studies of the kinematics, including infall toward the central regions, core rotation, and ambipolar diffusion (see, e.g., \citealt{arzoumanian26}) using multi-line observations that include lines at 3mm.}
\item{\textbf{High spatial resolution} ($\sim 500\,$au) to resolve their central regions both in terms of chemical segregation and dynamics of the gas, and study its connection to the outer envelope of the core.}
\item{\textbf{Sensitivity to large scales}: prestellar cores are extended ($>>1000\,$au), contracting objects, where matter keeps flowing from the larger cloud scales to the central zones.}
\end{enumerate}
\vspace{-4pt}
Currently there is no facility that allows such observations. The community initiated preliminary studies with the APEX telescope, which can observe all the target lines. However, APEX does not offer the requested angular (see Fig.~1) and frequency resolution, or the sensitivity to reach the goals presented above in a statistical way. Interferometers as ALMA, on the other hand, lack the large-scale sensitivity which is crucial to avoid missing flux problems in the solar neighbourhood (hence addressing sources that are representative of Solar-like systems; \citealt{tokuda25}). These observational requirements are all met by the technological advancements proposed by the Atacama Large Aperture Submillimeter Telescope (AtLAST) design study. In particular:
\vspace{-4pt}
\begin{enumerate}
  \item Located at 5000$\,$m near ALMA, AtLAST will operate from $\sim$ 30 to 950 GHz (0.3-10$\,$mm);
    \item Its high-frequency resolution backends will provide high-velocity resolution observations across the entire covered wavelength range;
\item The requirement of a $\sim 500\, $au resolution for nearby star-forming regions ($\sim$150$\,$pc) even at the lower frequencies ($1\,$mm) translate into at least a 50$\,$m dish, as the one proposed by AtLAST;
    \item the design study foresees a large 2$^\circ$ FoV, which
enables sensitive observations across large scale structures such as prestellar cores, allowing to recover faint, extended low surface brightness structures that are missed by interferometers, and enabling statistical studies by observing large ($\sim 50$ objects) samples.  
  
\end{enumerate}
A telescope with the above mentioned characteristics would additionally be revolutionary to address synergetic aspects of star and planet formation (e.g., dust growth and polarisation studies) that require sensitivity, large frequency coverage, and high frequency resolution backends, among other things, that are currently not available.

\vspace{-10mm}

\newpage
\bibliographystyle{abbrvnat}
\bibliography{bib.bib}

\end{document}